\documentclass[aps,prl,twocolumn,superscriptaddress,showpacs]{revtex4}

\usepackage[utf8]{inputenc}
\usepackage[english]{babel}
\usepackage{amsmath}
\usepackage{amssymb}
\usepackage{color}
\usepackage{bbold}
\usepackage[abs]{overpic}

\usepackage{graphicx}
\usepackage{cancel}

\newcommand{\eS}{\mathcal{S}_{\text{VN}}}

\newcommand{\eM}{\mathcal{M}}

\begin{document}

\title{Full characterization of the quantum linear-zigzag transition in atomic chains}

\author{Pietro Silvi}
\affiliation{Institut f\"ur Quanteninformationsverarbeitung,
  		Universit\"at Ulm, D-89069 Ulm, Germany}
\author{Gabriele De Chiara}
\affiliation {Centre for Theoretical Atomic, Molecular and Optical Physics,
		Queen's University of Belfast, Belfast BT7 1NN, United Kingdom}
\author{Tommaso Calarco}
\affiliation{Institut f\"ur Quanteninformationsverarbeitung,
  		Universit\"at Ulm, D-89069 Ulm, Germany}
\author{Giovanna Morigi}
\affiliation {Theoretische Physik, Universit\"at des Saarlandes,
		D-66123 Saarbr\"ucken, Germany}
\author{Simone Montangero}
\affiliation{Institut f\"ur Quanteninformationsverarbeitung,
  		Universit\"at Ulm, D-89069 Ulm, Germany}

\date{\today}

\begin{abstract}


A string of repulsively interacting particles exhibits a phase transition to a zigzag structure,
by reducing the transverse trap potential or the interparticle distance. The transition is driven by transverse,
short wavelength vibrational modes. Based on the emergent symmetry $\mathbb{Z}_2$ it has been argued that this instability
is a quantum phase transition, which can be mapped to an Ising model in transverse field. We perform an  extensive
Density Matrix Renormalization Group analysis of the behaviour at criticality and evaluate the critical exponents and the central
charge with high precision. We thus provide strong numerical evidence  confirming that the quantum linear-zigzag transition
belongs to the critical Ising model universality class. These results show that structural instabilities of one-dimensional
interacting atomic arrays can simulate quantum critical phenomena typical of ferromagnetic systems.
  
\end{abstract}


\maketitle

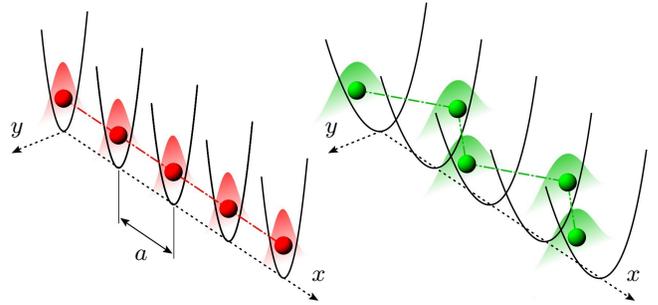
\begin{figure}
 \begin{center}
 \begin{overpic}[width = \columnwidth, unit=1pt]{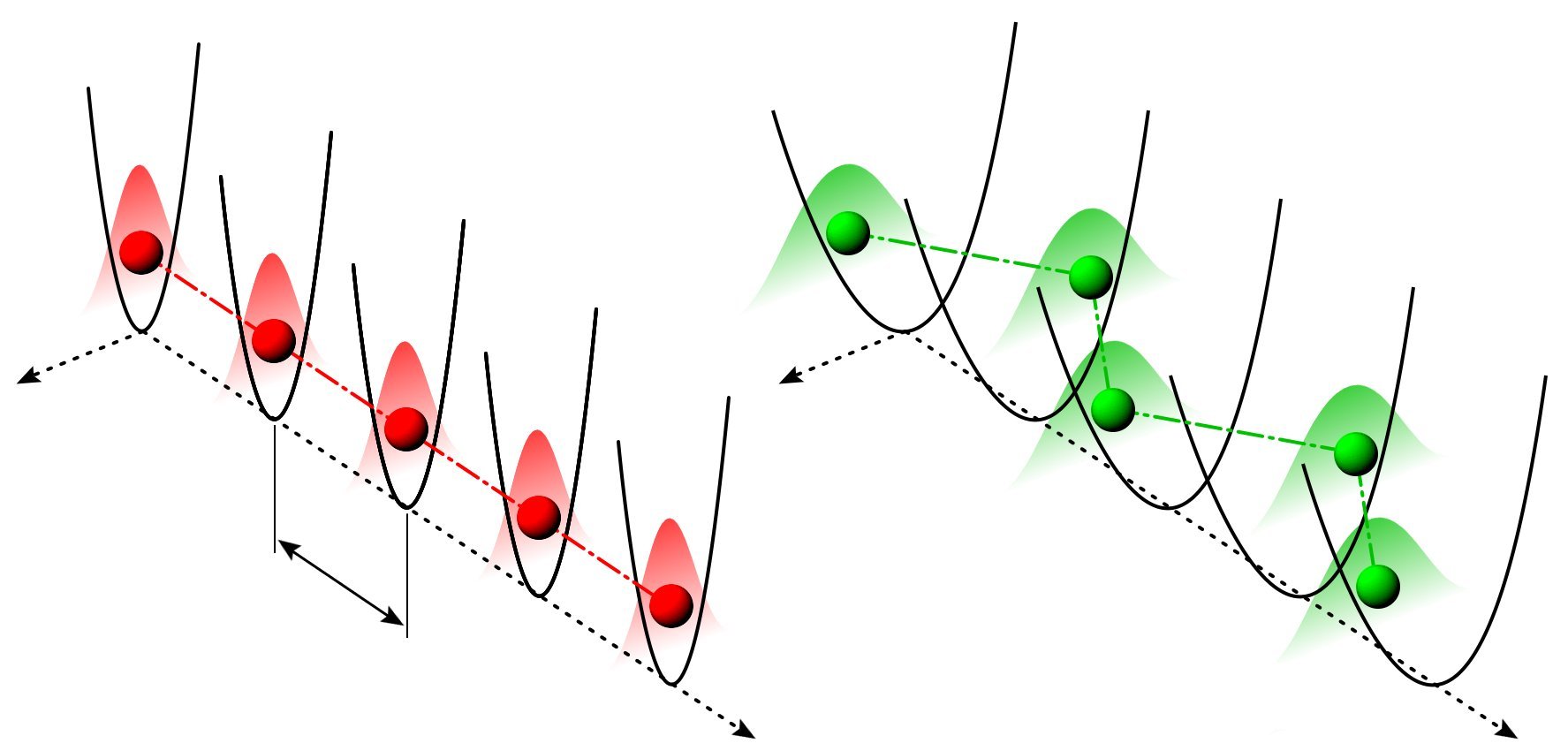}
  \put(2, 65){$y$}
  \put(121, 65){$y$}
  \put(116, 9){$x$}
  \put(235, 9){$x$}
  \put(49, 17){$a$}
 \end{overpic}
 \end{center}
\caption{ \label{fig:design} (color online)
Linear-zigzag instability in a chain of interacting atoms: (left) linear and (right) zigzag configuration. The transition is either controlled by changing the longitudinal lattice spacing $a$ or the frequency $\omega_T$ of the transverse harmonic confinement \cite{Morigi:G.5}. }
\end{figure}

Quantum simulators \cite{Bloch:1} are acquiring increasing prominence in the area of quantum technologies. The basic idea, dating back to Feynman \cite{Feynman}, is to create a controllable system whose dynamics reproduces that of a many-body quantum mechanical model that is both important and very difficult to solve. Experimental realisation in systems with high level of control, such as ultra cold atoms and ions, allows the study of the properties of the simulated model in a way otherwise very challenging for classical simulations  \cite{Lewenstein,Bloch:2,Schneider,PorrasCirac}. One prominent example is the simulation of quantum critical phenomena described by $\phi^4$ type of models \cite{KZatoms,KZions,Higgs:HH,Higgs:Demler}, which are encountered in solid state and high-energy models and whose predictions need verification \cite{Weinberg,Sachdev}.
In this context, the zigzag instability of interacting atomic chains is a realization of $\phi^4$ model which can be realized in laboratory \cite{Morigi:G.5,Retzker:2008,pyka12,Schaetz}. This  transition is sketched in Fig.~\ref{fig:design} and has been observed in systems of singly-charged ions confined by external potentials, where the instability is controlled either by lowering the transverse potential or the linear density \cite{zig1,Paul1}.  Due to its universal properties, it can be observed in arrays of other repulsively interacting (quasi) particles \cite{Shimshoni:2011a,Shimshoni:2011b}, such as electrons in quantum wires \cite{Meyer}, ultracold gases in optical lattices mutually repelling via the dipolar interaction \cite{Kollath,Morigi:Dipoli} or via the off-resonant coupling with a Rydberg excitation \cite{Pohl}, or vortices in quantum gases \cite{Busch}. In these systems ultralow temperatures are typically achieved, so that quantum effects at criticality are dominant. 
It has been argued that the linear-zigzag structural instability is a quantum phase transition~\cite{Shimshoni:2011a}, which in two dimensions can be mapped to an Ising model in the transverse field, describing a ferromagnetic  transition at zero temperature \cite{Sachdev}. This mapping was first proposed for Wigner crystals of electrons in quantum wires \cite{Meyer}, and then put forward in Ref.~\cite{Shimshoni:2011a}  using conformal-field-theory considerations motivated by the emergent $\mathbb{Z}_2$ symmetry.
These studies call for a precise numerical verification ruling out other possible effects,
which cannot be systematically accounted for when performing the mapping with analytic tools.

In this Letter we perform an extensive Density Matrix Renormalization Group (DMRG) study~\cite{White92, PrimoMPS, iblisdir07} and demonstrate that the linear-zigzag instability belongs to the universality class of the Ising model in a transverse field.  In particular, we quantify the quantum corrections to the classical linear-zigzag transition and relate their magnitude to experimental parameters. We remark that the 1D lattice $\phi^4$ model has been previously numerically investigated by means of other numerical methods~\cite{Barma,MontecarloP4}, including DMRG \cite{Sugihara}. Our work provides 
high precision values of the critical exponents and of the central charge of the system, giving for the first time irrefutable evidence that the discrete, quantum $\phi^4$ model in one dimension belongs to the universality class of the Ising model with transverse field. The computed critical exponents are summarized  in the following table together with the values predicted \cite{Sachdev}:
\[
 \begin{array}{|cl|c|c|}
  \hline
 & \mbox{\emph{Quantity}} & \mbox{\emph{Computed}} & \mbox{\emph{Theory \cite{Sachdev}}} \\
\hline
  \;\;\eta\;\; & \mbox{Anomalous dimension} & 0.258 \pm 0.012 & 0.25\\
  \beta & \mbox{Spont. magnetization} & 0.126 \pm 0.011 & 0.125 \\
  \nu & \mbox{Correlation length} & 1.03 \pm 0.05 & 1 \\
  c & \mbox{Central charge} & 0.487 \pm 0.015 & 0.5 \\
   \hline
 \end{array}
\]

{\it Model --} DMRG program simulates the quantum critical behaviour of a chain
of repulsively interacting atoms with mass $M$,
in presence of an anisotropic potential confining the motion in the $x-y$ plane,
by diagonalizing the Hamiltonian describing the dynamics. The atoms are assumed to form a regular array along the $x$ axis composed by $L$ sites with interparticle distance $a$. Quantum degeneracy effects are negligible due to either the strength of the interaction (for ions) or the external confining potential (e.g., a deep optical lattice for ultracold neutral atoms) \cite{Footnote:electrons}. The atoms position and canonically conjugated momentum are $(x_j,y_j)$ and $(p_{x,j},p_{y,j})$, with $j=1,\ldots,L$, such that $[x_j,p_{x,\ell}] = [y_j,p_{y,\ell}]=i\hbar \,\delta_{j,\ell}$. The Hamiltonian reads
 \begin{multline} \label{eq:original}
 H = \sum_{i = 1}^{L}
 \left[ \frac{p_{x,i}^2 + p_{y,i}^2}{2M} + \frac{M \omega_T^2}{2} y_i^2 + V_{\ell}(x_i)\right.\\ 
 \left. +
 \frac{1}{2} C_{\rm int} \sum_{j \neq i}
 \frac{1}{[(x_i - x_j)^2 + (y_i - y_j)^2]^{\alpha/2}} \right]\;,
\end{multline}
where $\omega_T$ is the frequency of the harmonic potential in the $y$-direction, $V_\ell(x)$ denotes the longitudinal confinement with a quasi-flat bulk and hard edges,
$C_{\rm int}$ gives the strength of the repulsive interaction. 
For trapped ions with same charge $Q$, $\alpha=1$, $C_{\rm int}=Q^2/(4\pi\epsilon_0)$,
and $a$ is of the order of micrometers \cite{zig1,Paul1}.
For ultracold dipolar gases with dipolar moment aligned perpendicular to the plane,
$\alpha=3$ and $C_{\rm int}=C_{dd}/(4\pi)$ with $C_{dd}$ the dipolar interaction strength \cite{Morigi:Dipoli}, and the interparticle distance is determined by the periodicity of an optical lattice along $x$ that traps the particles deep in the Mott-insulator phase \cite{Bloch:2,Menotti}. 

The diagonalization of Hamiltonian \eqref{eq:original} is a problem of high numerical complexity. Since the analysis is focused on the quantum ground-state properties and on the low energy excitations close to the zigzag instability, one can map Eq. \eqref{eq:original} to an effective one-dimensional Hamiltonian $H_{\rm eff}$, describing a short-range theory for the transverse motion \cite{Morigi:G.5,DelCampo,Shimshoni:2011a,Shimshoni:2011b}. It is convenient to report  $H_{\rm eff}$ in the dimensionless form $\tilde{H}=H_{\rm eff}/{\cal E}_0$, where ${\cal E}_0=C_{\rm int}/a^{\alpha}$ is a scalar with the dimension of an energy, while $\tilde{H}$ reads
\begin{multline} \label{eq:effective}
 \tilde{H} = \frac{1}{2} \sum_{i = 1}^{L}
 \left[ \tilde{p}_i^2 +
 \left( \tilde{\omega}^2 -
 \mathcal{M}_1 
   \right) \tilde{y}_i^2
 \right.  \\ + \left.
  \mathcal{M}_2  \,(\tilde{y}_i + \tilde{y}_{i+1})^2
 +
   \mathcal{M}_3  \; \tilde{y}_i^4
  \right] \;,
\end{multline}
and is a discretized, anti-ferromagnetic version of the well-known $\phi^4$-field-theory model~\cite{Sugihara, MontecarloP4}. Now, all physical quantities are dimensionless: $\tilde{y} = y/a$, $\tilde{p} = p /\sqrt{M{\cal E}_0}$, and $\tilde{\omega} = \omega_T/\sqrt{{\cal E}_0/(Ma^2)}$. The values $\mathcal{M}_{j=1,2,3}$ are constants, their generic dependence as a function of $\alpha$ and $C_{\rm int}$ is given in Ref.~\cite{Shimshoni:2011b}. For instance, for ions $\mathcal{M}_1 = \frac{7}{2}\,\zeta(3)$,  $ \mathcal{M}_2 = \ln 2$,   and $ \mathcal{M}_3 =  \frac{93}{8}\,\zeta(5)$, while for dipoles $\eM'_1 = \frac{93}{8}\, \zeta(5)$,  $\eM'_2 = \frac{9}{4}\, \zeta(3)$ and $\eM'_3 = \frac{1905}{32}\, \zeta(7)$, where $\zeta(n)$ is the Euler-Riemann zeta function. The rescaled space-momentum commutator reads $[\tilde{y}_j, \tilde{p}_{\ell}] \equiv ig\delta_{j,\ell}$ with 
\begin{equation} \label{eq:gidef}
g \equiv \sqrt{\frac{\hbar^2}{M a^2{\cal E}_0}}\,.
\end{equation}
The parameter $g$ replaces the Planck constant~$\hbar$ in the definition of Hamiltonian \eqref{eq:effective}. 
Hamiltonian $\tilde{H}$ is obtained from Eq. \eqref{eq:original} in the thermodynamical limit, keeping $a$ constant, and assuming that the transverse displacements are smaller than the typical lattice constant, $\sqrt{\langle y_j^2\rangle} \ll a$. In this regime, close to the zigzag instability the transverse motion is effectively decoupled from the longitudinal excitations~\cite{Morigi:G.5,DelCampo}. Moreover, in this limit the low frequency part of the dispersion relation of Hamiltonian Eq. \eqref{eq:original} is equivalent to the one of Eq. \eqref{eq:effective}, as shown in Refs. \cite{Morigi:G.5,Shimshoni:2011a,Shimshoni:2011b}.  

The rescaled Hamiltonian description of Eq.~\eqref{eq:effective} shows that the emergent short-range theory  depends only on two parameters: the rescaled transverse trap frequency $\tilde{\omega}$ and the the ``effective Planck constant'' $g$. The latter parameter is the square root of the ratio of the kinetic over the interaction energy, and thus it measures the strength of quantum fluctuations at criticality.  Landau theory is found when $g=0$: in this limit the critical value of the transverse frequency $\tilde{\omega}_c(0)$ is given by the equation $\tilde{\omega}_c(0)^2 = \mathcal{M}_1$, for which the quadratic on-site potential in Eq. \eqref{eq:effective} vanishes \cite{Morigi:G.5,Morigi:G.7}. When quantum fluctuations are relevant, tunneling between the two wells shifts the critical frequency to smaller values $\tilde{\omega}_c(g)\le\tilde{\omega}_c(0)$. This shift has been estimated in 
Ref.~\cite{Shimshoni:2011a,Shimshoni:2011b}. In this work, among other results, we provide a precise numerical determination of $\tilde{\omega}_c$. 

\begin{figure}
 \begin{center}
 \begin{overpic}[width = \columnwidth, unit=1pt]{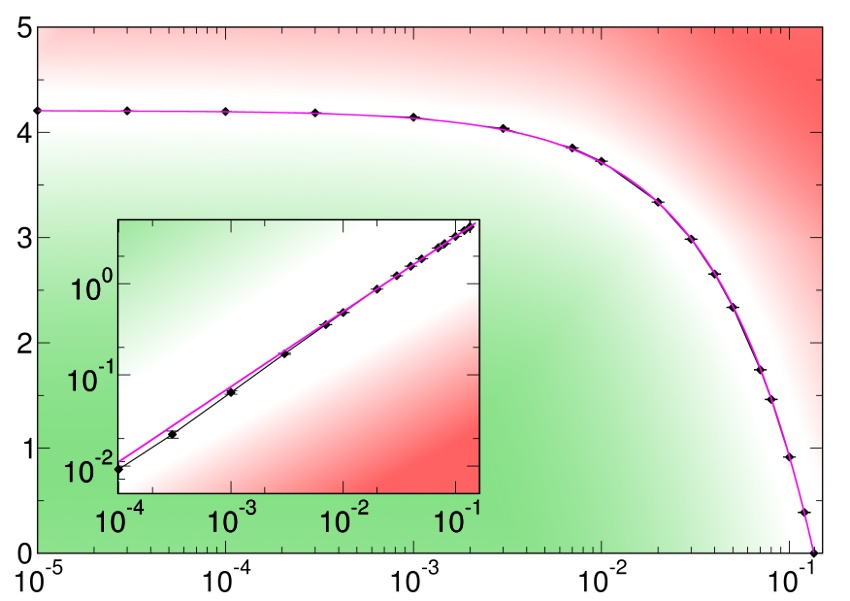}
  \put(-5, 85){\large $\tilde{\omega}^2$}
  \put(197, 0){\large $g$}
  \put(25, 115){$|\Delta \tilde{\omega}_c^2 |$}
 \end{overpic}
 \end{center}
\caption{ \label{fig:phasedia} (color online)
Phase diagram of an array of ions in the $(g, \tilde{\omega}^2)$ parameter space; the red (green) region
locates the linear (zigzag) phase. Inset: displacement of the critical square frequency $|\Delta \tilde{\omega}^2_c| $ as a function of $g$.
The magenta line is the power-law fit described by Eq.~\eqref{eq:omegagi}:
$| \Delta \tilde{\omega}_c^2 |= A g^r$, with $r = 0.82 \pm 0.001$ and $A = 21.9 \pm 0.1$. The phase diagram for dipoles, for instance, can be obtained via the rescaling $ {\tilde{\omega}}'^2 = \eM'_1 + (\tilde{\omega}^2 - \eM_1)\eM'_2/\eM_2 $ and $ g' = g \sqrt{\eM_3^2 \;\eM'^3_2/(\eM_2^{3} \;\eM'^2_3)}$, see text.
}
\end{figure}

To perform the DMRG simulations we introduce a local (finite) basis at every lattice site to reduce the 
continuous local variable $\tilde{y}_j$ into a discrete algebra suitable to fit into DMRG analysis~\cite{iblisdir07}. 
We solve the local part of the Hamiltonian $\tilde{H}$ in Eq. \eqref{eq:effective},
which is a	 size-independent,  homogeneous problem.
We numerically diagonalize the local Hamiltonian, defining an (infinite)
set of eigenfunctions and we truncate the local bases retaining only the $d$ lowest-energy states,
which are used to expand the full many-body Hamiltonian.
We check a posteriori the validity of the approximation 
we introduce by keeping track of the populations of the single-site reduced density matrices.
Typically, the matrix diagonal elements decay exponentially fast
with the level index, dropping below $10^{-5}$ already at $d \sim 8$.
We therefore employ a local basis dimension around $10<d<20$. We also typically adopt a DMRG bondlink
dimension up to $m \sim 30$ so that errors are kept well under control.
In this setup, we handle simulations up to $L \sim 3000$ sites with open boundary conditions.

{\it Results --}  We first characterize the phase diagram of the linear-zigzag transition at the thermodynamic limit in the 
$\tilde{\omega}$ and $g$ plane. The textbook order parameter for distinguishing between these phases
is the antiferromagnetic order parameter $\xi = \lim_{L \to \infty} L^{-1} \sum_{j=1}^{L} (-1)^{j} \tilde{y}_j$. 
We have only direct access to finite-size systems where we can measure the single site displacement 
$\langle \tilde{y}_j \rangle$ and then perform the limit. However, as there can be no spontaneous symmetry
breaking in finite-size samples, $\langle \tilde{y}_j \rangle = 0$ for every system size $L$. Indeed,  
the two ordered-phase configurations interfere and the system ground state is their even superposition. 
To overcome this problem we compute the order parameter indirectly from the two-point correlations
$\langle \tilde{y}_{j} \,\tilde{y}_{\ell} \rangle$, which are insensitive to the
symmetry breaking. More precisely, we compute the {square root of the structure factor density}:
\begin{equation} \label{eq:ordparam}
 \xi_L (g, \tilde{\omega}) = \sqrt{\frac{1}{L^2}  \sum_{j,\ell = 1}^{L} e^{i \pi (\ell - j)}
 \langle \Psi^{L}_{g, \tilde{\omega}} |
 \tilde{y}_{j} \cdot \tilde{y}_{\ell} 
 | \Psi^{L}_{g, \tilde{\omega}} \rangle },
\end{equation}
where $| \Psi^{L}_{g, \tilde{\omega}} \rangle$ is the ground state at $(g, \tilde{\omega})$ and size $L$.
Due to the sub-extensivity of entanglement-based quantum correlations, this definition coincides with the previous one at the thermodynamical limit. The phase diagram is determined by extrapolating  the order parameter
in Eq.~\eqref{eq:ordparam} at the thermodynamical limit: we estimate $\xi_\infty (g, \tilde{\omega}) \equiv \lim_{L \to \infty} \xi_L (g, \tilde{\omega})$ in order
to discriminate whether the $(g, \tilde{\omega})$ point lies in the linear ($\xi_\infty = 0$)
or in the zigzag phase ($\xi_\infty > 0$). Figure \ref{fig:phasedia} shows the resulting phase diagram.
The critical boundary appears to follow a power-law behavior, a numerical fit of the displacement from the classical critical point  results in the formula
\begin{equation} \label{eq:omegagi}
\Delta \tilde \omega_c^2 = \tilde{\omega}_c^2(g) - \tilde{\omega}_c^2(0)=- (21.9 \pm 0.1) \cdot g^{0.82 \pm 0.01}.
\end{equation}

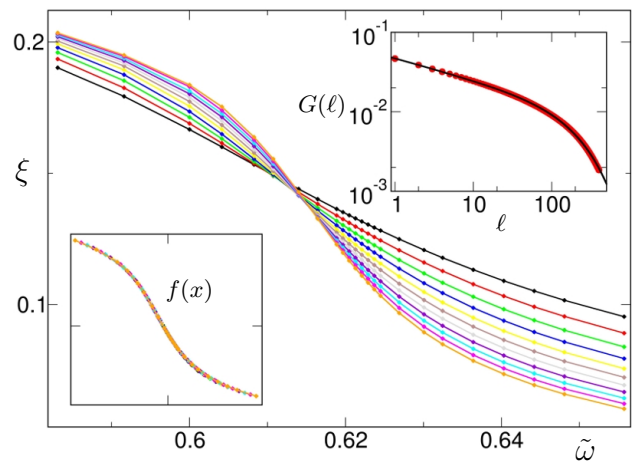
\begin{figure}
 \begin{center}
 \begin{overpic}[width = \columnwidth, unit=1pt]{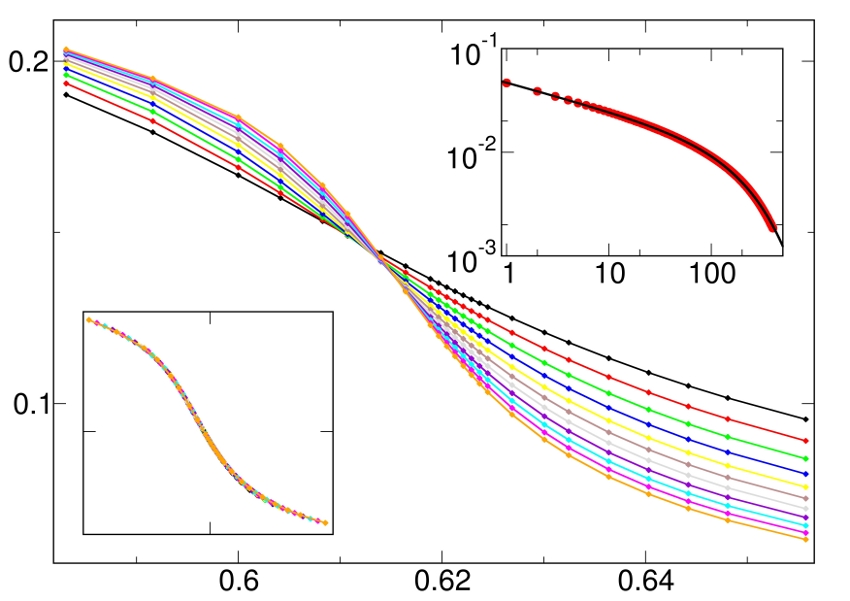}
  \put(3, 104){\large $\xi$}
  \put(215, -2){\large $\tilde{\omega}$}
  \put(60, 60){$f(x)$}
  \put(185, 85){$\ell$}
  \put(110, 130){$G(\ell)$}
 \end{overpic}
 \end{center}
\caption{ \label{fig:finisize} (color online)
Order parameter $\xi$ as a function of $\tilde{\omega}$, for different $L = 100, 120 \ldots 300$.
Bottom-left inset: rescaled data according to $\xi_L (\tilde{\omega})$, characterizing $f(x)$ (see text).
Top-right inset: two points correlation $\langle \tilde{y}_j \tilde{y}_{j+\ell} \rangle$ (red dots)
as a function of the distance $\ell$ (the black line shows a quasi-critical fit, see text).
}
\end{figure}

To further characterize the quantum phase transition we numerically evaluate the critical exponents. 
In particular, the transverse displacement $\tilde{y}$ plays the role of the direction of the spontaneous magnetization
in the Ising model, thus the decay rates of correlators $\langle \tilde{y}_j \tilde{y}_{j+\ell} \rangle$ as a function of the distance $\ell$
reveal the \emph{anomalous dimension} critical exponent $\eta$. To extract this quantity, we fit $\langle \tilde{y}_j \tilde{y}_{j+\ell} \rangle$ far from the boundaries using equation 
$ \langle \tilde{y}_j \tilde{y}_{j+\ell} \rangle \simeq G(\ell) = \alpha \,\ell^{-\eta} \,\exp\left( - \ell / \lambda \right)$. 
The exponential correction to the power-law decay compensates for not being exactly 
at the critical point by introducing a finite correlation length $\lambda$~\cite{Mussardo}.
As it can be seen in the topmost inset of Fig.~\ref{fig:finisize},
the agreement is almost perfect over lengths of hundreds of lattice sites.
We averaged the fitted exponent $\eta$ for different values of $g$ while being as close as possible to the phase boundary,
i.e.~maximizing $\lambda$. The final result is $\eta_{\text{av}} \simeq 0.258 \pm 0.012$, 
in good agreement with the predicted value of $1/4$ \cite{Sachdev}.

The spontaneous magnetization $\beta$ and correlation length divergence $\nu$ are
extracted through a finite-size scaling of the order parameter $\xi_L (\tilde{\omega})$ defined in Eq. \eqref{eq:ordparam}.
Following a procedure based on renormalization group analysis,
we find that near the transition the order parameter follows the scaling 
$ \xi_L (\tilde{\omega}) \simeq L^{-\beta / \nu} \; f \left( (\tilde{\omega} - \tilde{\omega}_c) \cdot L^{1 / \nu} \right), $
for some non-universal function $f(x)$ which may be model dependent \cite{FishBarb}.
It is then
possible to plot $L^{\gamma_1} \xi_L$ as a function of $(\tilde{\omega} - \tilde{\omega}_c) \; L^{\gamma_2}$, 
and then tune the exponents $\gamma_1$ and $\gamma_2$
to collapse all the curves onto one another. 
A typical example of such analysis is depicted in Fig.~\ref{fig:finisize}.
As a result of this rescaling we obtain $\beta_{\text{ext}} = {\gamma_1}/{\gamma_2} \simeq 0.126 \pm 0.011$ and
$\nu_{\text{ext}} = {1}/{\gamma_2} \simeq 1.03 \pm 0.05$ in perfect agreement with the theory \cite{Sachdev},
which predicts $1/8$ and $1$ respectively. This finite-size scaling procedure gives also another estimate 
of the critical boundary, in perfect agreement with the results showed in Fig.~\ref{fig:phasedia}.

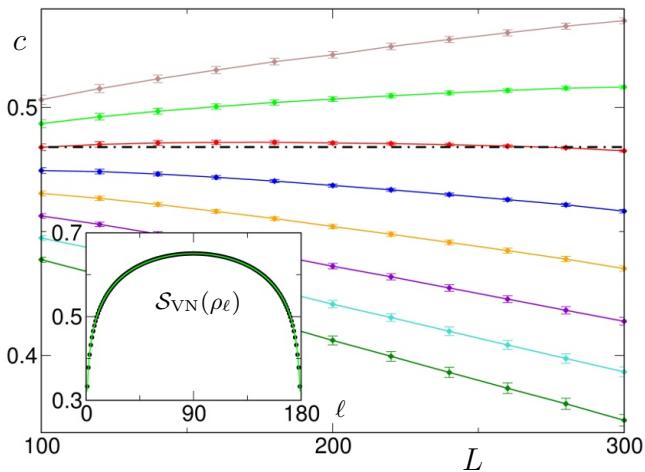
\begin{figure}
 \begin{center}
 \begin{overpic}[width = \columnwidth, unit=1pt]{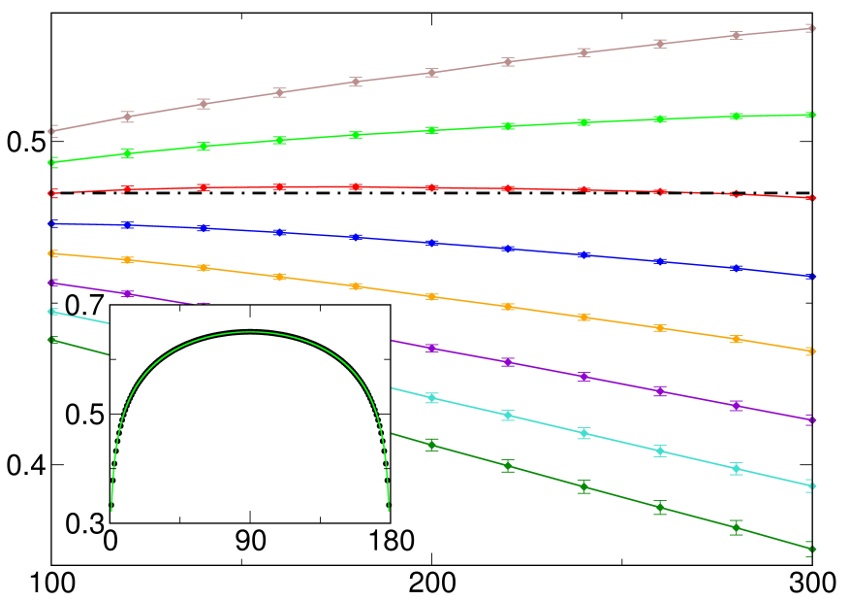}
  \put(4, 155){\large $c$}
  \put(174, -4){\large $L$}
  \put(126, 16){$\ell$}
  \put(58, 57){$\eS(\rho_\ell)$}
 \end{overpic}
 \end{center}
\caption{ \label{fig:central}
Inset: Von Neumann entropy $\eS(\rho_{\ell})$ (black dots) as a function of the partition size $\ell$
fitted via $c$, $c'$ using equation \eqref{eq:cardy} (green line).
Main: $c$ values as a function of $L = 100,120 \ldots 300$
for different $\omega = 0.383, 0.384 \ldots 0.390$ (top to bottom) at $g = 0.12$ for
various total lengths $L$ and trap frequencies $\tilde{\omega}$ (here at $g = 0.12$).
At the critical point $\tilde{\omega}_c$ (red data set), $c$ is constant with $L$ 
and it estimates the central charge.
}
\end{figure}

Finally, we measure the central charge $c$ of the model~\cite{Cencharge}. 
It is actually possible to extract the central charge of a critical ground state by computing the entanglement related to a left-right
($1..\ell \leftrightarrow \ell+1 .. L$) system partition~\cite{VonNeu}. That is, we evaluate the Von Neumann Entropy $\eS$ of the reduced density matrix $\rho_\ell$ of
either partition:
$\eS(\rho_{\ell}) \equiv - \mbox{Tr} \left[ \rho_{\ell} \ln \rho_{\ell} \right]$.
It is straightforward to determine the partition entropy when the quantum state is obtained via DMRG~\cite{PrimoMPS, Arealaws}. In fact, DMRG algorithms provide directly the Schmidt coefficients $\mu_p(\ell)$ for every partition $\ell \in \{1..L\}$,
from which the Von Neumann entropy $\eS(\rho_{\ell}) = - \sum_p \mu^2_p(\ell) \ln \mu^2_p(\ell)$ is easily computed.
The profile $\eS(\rho_{\ell})$ at the critical point grows as
\begin{equation} \label{eq:cardy}
  \eS(\rho_\ell) = \frac{c}{6} \log \left(L \cdot \sin{\frac{\pi \ell}{L}} \right) + c'\;,
\end{equation}
which allows to directly fit the central charge of the system~\cite{Calacardy, dechiara06}.
Figure~\ref{fig:central} displays a typical result of this analysis: for every system size $L$ and 
trapping frequency $\tilde \omega $ we fit the central charge $c$ by using Eq.~\eqref{eq:cardy} (inset).
We then plot  the fitted values of the central charge as a function of the system size $L$ and extract the value 
that is size-independent, and thus critical. We average over different values of $g$ and obtain $c_{\text{av}} \simeq 0.487 \pm 0.015$, which is in good agreement with the predicted value of $1/2$ \cite{Sachdev}.

In experiments the quantum critical behaviour can be measured, for instance, via the structure form factor \cite{Shimshoni:2011b} or by quenches across the critical point~\cite{Grandi}. The quantum disordered region is accessed for temperatures $T$ that are smaller than the energy scale of the gap of the Ising model. For detailed discussions we refer the reader to Ref. \cite{Shimshoni:2011a}. Our calculation allows us to precisely determine the frequency width of the disordered phase. This reads $\Delta\omega_c=\sqrt{{\cal E}_0/(Ma^2)}\Delta \tilde \omega_c$, where $\Delta \tilde \omega_c$ is given in Eq.~\eqref{eq:omegagi} as a function of the parameter $g$. For ions $g$ is proportional to the linear density $1/a$, which is limited by the Coulomb repulsion, and takes values between $10^{-5}$ and $10^{-4}$ giving a small critical region \cite{Shimshoni:2011a}. For dipolar gases and Rydberg-dressed gases  the situation can be quite different \cite{Morigi:Dipoli,Pohl}. For dipoles, for instance, $g=\sqrt{a/r_
0}$,  where $r_0=M C_{\rm int}/\hbar^2$ is the characteristic length of quantum coherence, $r_0\lesssim 100\mu$m, while long-range order can be realized by means of an optical lattice (with the molecules deep in the Mott-insulator phase, so that quantum fluctuations along the array can be neglected~\cite{Menotti}) so that $g > 0.01$. Note that, in the absence of an external periodic lattice, the linear-zigzag transition in quantum gases can be observed at rather large densities \cite{Morigi:Dipoli,Altman}. 

To resume, we implemented a DMRG program which allowed us to characterize the ground state of a one-dimensional array of interacting atoms close to the linear-zigzag transition. The universality class of the criticality was identified by extrapolating the  critical exponents as well as the central charge of the quantum phase transition.  The excellent match between the predicted values and the results of the simulations demonstrates the correspondence between linear-zigzag instability and the Ising universality class. These results show that one-dimensional quantum critical phenomena typical of ferromagnetic systems, recently simulated by means of engineered coupling between internal and external degrees of freedom of ions \cite{PorrasCirac,Schneider}, can be naturally simulated by structural instabilities of interacting atomic arrays. 

We acknowledge support from EU through PICC, AQUTE, and from the German Research Foundation (Heisenberg programme, SFB/TRR21), and the BW-grid for computational resources. We thank S. Fishman, A. Muramatsu, M.B. Plenio, A. Retzker, and E. Shimshoni for discussions,  and D. Rossini for contributing in developing the numerical code. PS acknowledges D. Fioretto, M. Burrello for stimulating discussions.


\begin{thebibliography}{1}
\bibitem{Footnote:electrons} For electrons the exchange interaction gives an additional term in Eq. \eqref{eq:effective} and thus to different critical regions \cite{Meyer}.
\end{thebibliography}
\end{document}